\documentclass[prl,aps,twocolumn,groupedaddress,showpacs]{revtex4}
\usepackage{amsmath}
\usepackage{amsbsy}
\usepackage{amssymb}
\usepackage[dvips]{graphicx}
\usepackage{epsfig}         

\newcommand{\e}{\epsilon}
\newcommand{\ms}{{m_s}}
\newcommand{\rav}{]_{av}}
\newcommand{\beq}{\begin{equation}}
\newcommand{\eeq}{\end{equation}}
\newcommand{\la}{\langle}
\newcommand{\ra}{\rangle}

\newcommand{\SG}{{\mbox{{\tiny G}}}}
\newcommand{\cop}{{\mbox{{\tiny}}}}

\begin{document}
\title{Phase diagram, correlation gap, and critical properties 
of the Coulomb glass}

\author{Martin Goethe}
\author{Matteo Palassini}\email{palassini@ub.edu}
\affiliation{Departament de F\'\i sica Fonamental,
Universitat de Barcelona, Diagonal 647, E--08028 Barcelona, Spain.}

\date{June 30, 2009}

\begin{abstract}
We investigate the lattice Coulomb glass model in 
three dimensions via Monte Carlo 
simulations. No evidence for an equilibrium glass phase is found 
down to very low temperatures, although the correlation length increases
rapidly near $T=0$. \hspace{-0.2cm} A charge-ordered phase (COP) exists 
 at low disorder. The transition to this phase is consistent with
the Random Field Ising universality class, which shows that the interaction 
is effectively screened at moderate temperature.
For large disorder, the single-particle density of states 
near the Coulomb gap satisfies the scaling relation 
$g(\e,T)=T^\delta f(|\e|/T)$ with  $\delta = 2.01\pm 0.05$
in agreement with the prediction of Efros and Shklovskii. 
For decreasing disorder, a crossover to a larger effective
exponent occurs due to the proximity of the COP.
\end{abstract}

\maketitle

In disordered insulators, the localized electrons cannot
screen effectively the Coulomb interaction at low temperature. Therefore,
many-electron correlations are important in this regime.
The long-range repulsion induces a soft ``Coulomb gap''
in the single-particle density of states (DOS).   
Efros and Shklovskii (ES) \cite{ES} argued that the gap
has  a universal form
$g(\e) \propto |\e-\mu|^{\delta}$ near the chemical
potential $\mu$, with  $\delta \geq d-1$ in $d$ dimensions,
and that a saturated bound $\delta=d-1$ modifies 
the variable-range hopping 
resistivity $\ln R \sim T^{-x}$ from Mott's law $x=1/(d+1)$ to
$x=1/2$.
Both the existence of the gap and the crossover to 
$x\simeq 1/2$   at low temperature $T$ have been confirmed experimentally
and in numerical simulations \cite{crossover}, but the validity of 
$\delta=2$ for $d=3$ has yet to be firmly established.
Pseudo ground-state numerical calculations gave
$\delta= 2.38$ \cite{li_phillips}, $\delta = 2.7$
\cite{moebius_dos,sarvestani,overlin}, $\delta \leq 2.01$ \cite{surer}, 
while finite-$T$ simulations obtain
$\delta$ between $2$ and $4.8$ \cite{li_phillips,sarvestani,overlin}
from the filling of the gap as
$g(\mu)\propto T^\delta$ \cite{levin,vjs,hunt}.

It was also suggested long ago \cite{davies}
that disordered insulators enter  
 a glass state at low temperature. 
Ample experimental and numerical evidence of glassy nonequilibrium
effects in these systems
has been obtained since \cite{see}. 
However, it remains unclear whether these effects are purely dynamical or reflect
an underlying transition to an {\em equilibrium} glass phase 
(GP), and whether there is a link between glassiness 
and the Coulomb gap.
Some evidence for a sharp equilibrium transition to
 a GP was found in simulations of localized charges
with random positions \cite{grannan,overlin,diaz}
but not in the presence of on-site disorder \cite{diaz,surer}.
In the latter case, the transition would not break any 
symmetry of the Hamiltonian, similarly to the long-debated
Almeida-Thouless  transition in spin glasses \cite{AT}.
These issues have been brought  again to the fore by recent 
mean-field studies \cite{pastor,pankov, ioffe, mueller} which predict
 a ``replica symmetry broken'' equilibrium GP below a critical temperature
$T_{\text{g}}$ in the presence of on-site disorder.
In this GP correlations remain critical, which leads to 
$\delta=d-1$, and both $T_{\text{g}}$  and
the gap width $\Delta$ scale as $W^{-\frac{1}{2}}$ 
for $d=3$ and large disorder strength $W$ \cite{mueller}.
\begin{figure}[h]
\hspace{-0.5cm}\includegraphics[height=0.84\linewidth,angle=270]{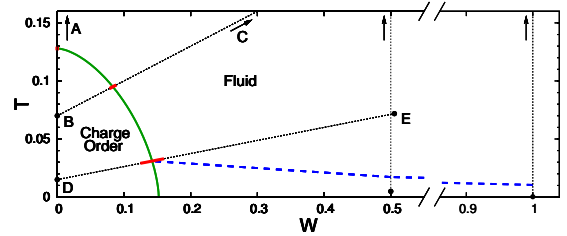}
\caption{(Color online) Phase diagram of the Coulomb glass model.
The thin lines are the simulation paths [BC: $T = (3/10) W + 7/100$, 
$W_{\text{C}}=0.977$; DE: 
$T=(9/80) W + 3/200$, $W_{\text{E}}=0.506$; $T_{\text{A}}=0.275$.]
The fluid-COP boundary interpolates
the transition temperatures estimated along AB, BC, DE, shown
in red.
Our results indicate that no glass phase exists above the
blue dashed line. See also Fig.2 of Ref.\cite{pankov}.}
\vspace{-0.3cm}
\end{figure}

In this Letter, we investigate these predictions
 via extensive Monte Carlo (MC) simulations
of the Coulomb glass lattice model with on-site disorder \cite{efros}.
In addition, we study in detail the transition from the fluid
to the charge-ordered phase (COP).
For $W=0$, there is good numerical  evidence for an Ising-like transition
\cite{moebius}.
For $W\neq 0$, mean-field theory predicts a stable COP
for $d=3$ \cite{pankov,malik}.
Beyond mean field there is some numerical evidence
that the COP survives small positional disorder
\cite{overlin,surer} and on-site disorder \cite{moebius_talk},
but neither the phase diagram nor the critical properties
have been investigated.  Our results are as follows:
(i) A COP exists below the (approximate) phase boundary in Fig.~1.
(ii) The fluid-COP transition is consistent
with the Random Field Ising model (RFIM) universality class,
which shows that the interaction is effectively screened
near the transition.
(iii) No  GP is found for $T$
well below the mean-field $T_{\text{g}}$
\cite{pankov}, in agreement
 with the results of Ref.~\cite{surer} but at lower
$T$ and in a wider range for $W$.
(iv) Due to the long-range interaction, the glass correlation length 
increases rapidly and possibly diverges
as $T\to 0$.
(v) For large $W$, the DOS scales as
$g(\epsilon,T)=T^\delta f(|\e|/T)$ near the gap, with a saturated
exponent $\delta\simeq 2.00$. (vi) As $W$ 
decreases, scaling breaks
down above the COP, and 
an effective power law $g_L(\e,T=0)\propto |\e|^\delta$ holds
with $\delta \gtrsim 3$, in contrast with Ref.\cite{surer}.
A more extended account will appear later \cite{gp}.

{\em Model and simulation --} We study
the Hamiltonian
\beq
\mathcal H =  \frac{e^2}{2 \kappa}\sum_{i\neq j}(n_i - K)\frac{1}{|{\bf r}_{ij}|}(n_j - K)
+ W \sum_i n_i \varphi_i
\label{ES}
\eeq
where $n_i \in \{0,1\}$ are 
the occupation numbers for the $N=L^d$ sites  
of a hypercubic lattice  ($d=3$)
with $\sum_{i=1}^N n_i = K N$,
and ${\bf r}_{ij}$ is the distance 
from $i$ to $j$. The filling factor is $K=1/2$ 
(which gives $\mu=0$).
The random on-site energies $\varphi_i$ are
independen and Gaussian-distributed with zero mean and variance unity.
Energies and temperatures will be
in units of $e^2/({\kappa} \ell)$ and lengths in units of the 
lattice spacing $\ell$.

We carry out canonical MC sampling 
along the paths ABC and 
ADE in Fig.~1 and at constant $W=0.2, 0.5, 1, 2, 4$.  We consider
an infinite sphere of periodic images of 
a central $L^3$ cell and sum over all interactions with the
Ewald method with a dipole surface term \cite{leeuw}.
To reach low temperatures,
we use the exchange MC algorithm \cite{exchange}. 
For each
realization (sample) $\varphi = \{\varphi_i\}_{i=1}^N$,
we simulate identical replicas with
different ($T,W$) along the simulation path.
Every $N/2$ Metropolis steps for single-electron hops,
replicas at adjacent $(T,W)$, $(T',W')$ swap
their configurations with 
probability $\min(1,p)$, where $p=\exp[(\beta - \beta')(\mathcal H-\mathcal H') + 
(W'-W)(\beta \mathcal R'-\beta' \mathcal R)]$,
$\beta=1/T$, and $\mathcal R = \sum_i n_i \varphi_i$, 
which preserves detailed balance.
The simulation time $t_{s}$ 
is chosen so that averages over the
intervals $[t_{s}/3, t_{s}]$ and   $[t_{s}/9, t_{s}/3]$ 
agree within the statistical errors, and that the 
identity 
$2 T N^{-1} [  \la \mathcal R\ra  \rav
= W (2 N^{-1} \sum_{i=1}^N [\la n_i^{(a)} n_i^{(b)} \ra \rav - 1)$,
valid for Gaussian disorder, is satisfied. Here,
$\la \cdot \ra$  and $[\cdot\rav$ are the thermal and sample
averages and $a,b$ are two independently simulated 
replicas with the same ($\varphi,T,W$) \cite{method}.

{\em Charge ordering --}
Fig.2 (top inset) shows the COP order parameter 
$M_s = [\la |m_s| \ra\rav$ along the paths AB, BC, and DE,
where  $m_s = N^{-1} \sum_{i=1}^N \sigma_i$
and  $\sigma_i = S_i (-1)^{x_i+y_i+z_i}$
(we introduce the Ising variables $S_i = 2 n_i -1$).
The sharp increase demonstrates a transition to a COP.
To determine the transition temperature
$T_{\text{c}}$, we measure 
the finite-size correlation length (CL) \cite{cooper}
\beq
\xi_{\tiny{L}} = \frac{1}{2 \sin (|{\bf k}_{\text{min}}|/2)} 
\left( \frac{\chi_L({\bf 0})}{\chi_L({\bf k}_{\text{min}})}-1 \right)^{1/2}\, ,
\label{xidef}
\eeq
where
$\chi_L({\bf k})= N^{-1} \sum_{i,j} [ \la \sigma_i \sigma_j \ra ]_{av} 
e^{i {\bf k} \cdot 
{\bf r}_{ij} }$
and ${\bf k}_{\text{min}} = (2\pi/L,0,0)$.
Along BC, the data for $\xi^{\cop}_L(T)/ L$ for different $L$ 
cross (Fig.2, main panel), which signals \cite{ballesteros2} a transition
at $T_{\text{c}}^{\text{BC}}= 0.0950(15)$. 
We observe similar crossings along AB and DE (not shown)
at $T_{\text{c}}^{\text{AB}} = 0.1280(15)$, in excellent agreement
with Refs.\cite{overlin,moebius}, and
$T_{\text{c}}^{\text{DE}} = 0.031(2)$.
The curve $(T_{\text{c}}(W)/T_{\text{c}}^{\text{AB}})^{1.60} = 1 -
(W/0.15)^{1.60}$ interpolates these three
points and gives 
the approximate fluid-COP phase boundary in Fig.~1.

\begin{figure}
\includegraphics[height=0.95\linewidth,angle=270]{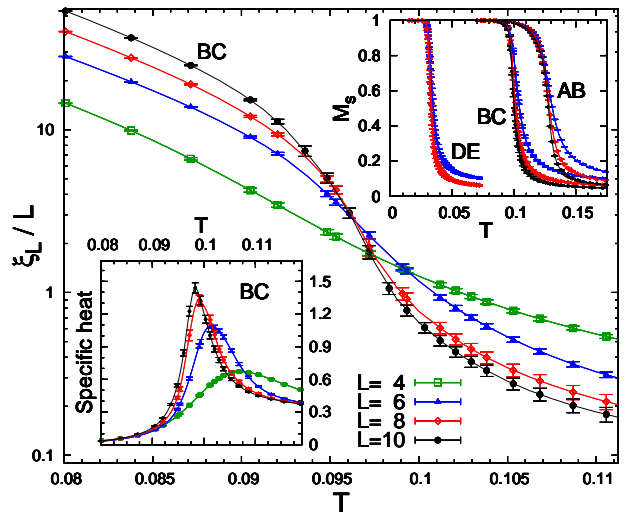}
\caption{(Color online) Charge-order CL 
along path BC in Fig.~1. {\bf Top inset}: order parameter $M_s$
along paths AB, BC, and DE. {\bf Bottom inset}: specific heat along path BC.}
\vspace{-0.3cm}
\end{figure}

{\em Critical behavior --}
Since  at $W=0$ the fluid-COP transition has
a positive specific-heat exponent \cite{moebius}, disorder is relevant 
and the $W\neq 0$ transition will be
governed by a random fixed point which, by analogy with the RFIM
\cite{moebius}, we expect to be at $T=0$ \cite{fisher}.
Assuming that the $W\neq 0$ transition is second order
(indeed the distribution of $m_s$ is unimodal at all $T$ for 
a predominant,  and increasing with $L$, fraction 
of the samples \cite{gp})
 we obtain the critical exponents in Table I.
$\beta/\nu$ and $\bar{\gamma}/\nu$ were estimated with
the {\em quotient method}\/ \cite{ballesteros} for
the observables 
$M_s$ and $\bar{\chi}^{\cop}_L= N [\la \ms^2 \ra \rav$
respectively [the quotient estimates from 
$(L,L')=(6,8),(6,10)$ and $(8,10)$ agree within the errors],
while $\gamma/\nu$ was obtained by fitting
$a L^{\gamma/\nu}$ to the height of the peak 
of the susceptibility
$N [\la \ms^2 \ra - \la |\ms| \ra^2 \rav$ (data not shown).
The peak height for 
the specific heat $c_L=1/(N T^2) [\la \mathcal{H}^2 \ra -\la
\mathcal{H}\ra^2\rav$ increases 
slowly with $L$ (Fig.~2, bottom inset),  which suggests either 
$\alpha<0$ or a logarithmic divergence ($\alpha=0$). 
We could not directly estimate $\nu$ in a reliable way, but
we obtain $\nu = 1.11(12)$ from
the modified hyperscaling relation  \cite{fisher} $(d - \theta)\nu = 2 - 
\alpha$, assuming $\alpha=0$  
and using $\theta = \bar{\gamma}/\nu - \gamma/\nu = 1.20(20)$.
As shown in Table I, the exponents agree 
fairly well with the known values for the RFIM \cite{middleton}, which
suggests that the interaction is effectively short-range 
near the phase boundary.
\begin{table}
\caption{
Critical exponents for the fluid-COP transition along BC in Fig.~1,
compared with  the RFIM values \cite{middleton}.}
\begin{tabular*}{\linewidth}{@{\extracolsep{\fill}} l l l l l l}

\hline
\hline

 & $\gamma/\nu\hphantom{2}$  &  $\bar{\gamma}/\nu\hphantom{2}$  & $\beta/\nu\hphantom{2}$ & $\nu\hphantom{2}$\\
\hline
Coulomb glass & $1.69(17)$ & $2.89(9) $  &$0.06(4) $ & $1.11(12)$ \\

RFIM          & $1.44(12)$ & $2.93(11)$  &$0.011(4)$ & $1.37(9) $  \\

\hline
\hline
\end{tabular*}
\vspace*{-0.50cm}
\end{table}

{\em Glass phase --}
Several works have searched for a GP by
measuring the parameter $[(\la n_i \ra -1/2)^2]_{av}$ \cite{davies}
or higher cumulants of the overlap between two replicas \cite{diaz}.
We measure instead the glass CL $\xi^{\SG}_{\tiny{L}}$
obtained from Eq.~(2) by replacing $[\la \sigma_i \sigma_j \ra]_{av}$
with the ``spin-glass'' correlation function $G({\bf r}_{ij})= 
[ (\la S_i S_j \ra - \la S_i \ra \la S_j \ra )^2 \rav$.
In the fluid phase we have
$G(r) \sim \exp(-r/\xi^{\SG})$ 
for  $\xi^{\SG} < r \ll L$,
where $\xi^{\SG}$ is the bulk CL,
thus 
$\xi^{\SG}_L \sim \xi^{\SG}$ for $L\gg \xi^{\SG}$
and $\xi^{\SG}_L \sim L$ for $L\ll \xi^{\SG}$.
In a ``many-state'' GP \cite{mueller},
$G(r)$ tends to a constant for large $r$, thus we have
$\xi^{\SG}_L \sim L^{d/2 + 1}$. Hence
the existence of a GP will be signaled by the crossing of 
$\xi^{\SG}_L(T)/ L$ 
for different $L$ near $T=T_{\text{g}}$ \cite{ballesteros2}.
\begin{figure}[h]
\includegraphics[height=\linewidth,angle=270]{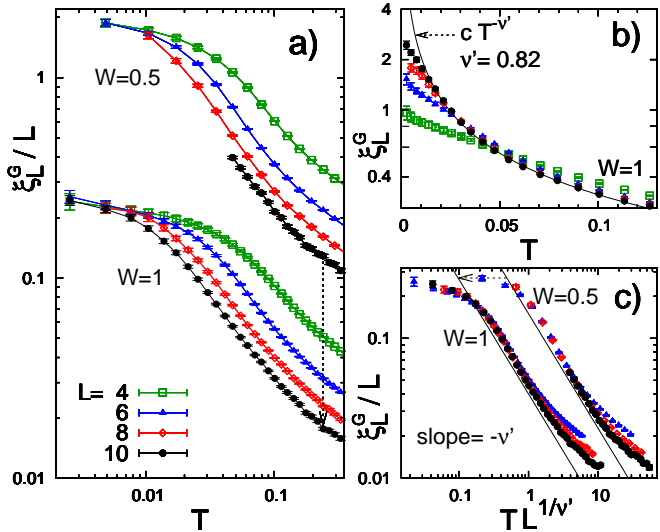}
\caption{(Color online) {\bf (a)} Glass correlation length $\xi^{\SG}_L$
at $W=1$ and $W=0.5$ (shifted upwards by a factor 7). The absence
of crossing is evidence against the existence of an equilibrium
glass transition. {\bf (b)} Power-law fit of $\xi_{10}^{\SG} = c T^{-\nu^\prime}$ 
for $W=1$. {\bf (c)} Scaling plot $\xi^{\SG}_L = L f(T L^{1/\nu^\prime})$ for
$W=1$ and $W=0.5$ (shifted to the right by a factor 5).}
\vspace{-0.3cm}
\end{figure}
As shown in Fig.3(a) for $W=0.5, 1$, we observe no crossing 
down to the lowest equilibrated temperature and well below the 
mean-field glass transition ($T_{\text{g}}\approx 0.037$ for $W=0.5$ \cite{pankov}).
Similar results 
were found in Ref.\cite{surer} for $T\geq 0.03$ and
$W\leq 0.4$.
We also exclude that a GP occurs at $T_{\text{g}}>T_{\text{c}}$ 
along BC and DE by comparing the crossing temperatures for
$\xi^{\SG}_L/L$ (not shown) and $\xi^{\cop}_L/L$:
For all pairs  $(L,L')$
they differ by less than $1 \%$. Together 
with the results at constant $W$, this indicates
that no GP exists above the dashed line in Fig.~1.

Fig.~3(b) shows that $\xi^{\SG}_L$ is nearly
independent of $L$
at large $T$ and $W=1$ apart from small finite-size effects,
thus $\xi^{\SG}_{L=10}$ is a good estimate of $\xi^{\SG}$ 
for $T\gtrsim 0.017$. 
At lower $T$ we observe $\xi^{\SG}_L\sim L$, which
shows that $\xi^{\SG}$ becomes 
larger than $\simeq 10$, and possibly diverges  as $T\to 0$.
Indeed, $\xi_{L=10}^{\SG}(T)$ can be fitted for $T\in[0.017,0.113]$  
by a power law $T^{-\nu^\prime}$ with $\nu^{\prime}=0.82$ [Fig.~3(b)], which
also gives a satisfactory finite-size scaling 
$\xi^{\SG}_L = L f(T L^{1/\nu^\prime})$ [Fig.~3(c)]. 
A divergence would be a nontrivial prediction,
since for $W\neq 0$ the ground state of a periodic sample is unique.
Because $L$ and $\xi^{\SG}_L$ are rather  small, however, we cannot rule out
neither that $\xi^{\SG}$ stays finite at $T=0$, nor
an exponent $\nu^{\prime}=1$. 
An interesting question is whether there is a link
with the $T^{-1}$ divergence of the screening length found
in mean-field theory \cite{mueller} and from 
simple arguments \cite{hunt,tf}.
We tested that the CL
obtained by replacing $[\la \sigma_i \sigma_j \ra]_{av}$ 
with $[\la \sigma_i \sigma_j \ra -  
\la \sigma_i \ra \la \sigma_j \ra]_{av}$ in 
Eq.~(2), remains smaller than unity
at all $T$, which suggests that the correlated regions are
disordered.
Finally, we simulated the short-range RFIM choosing $W$ so that
the value of $N^{-1}[\la \sum_i n_i \varphi_i\ra]_{av}$ is close to the
Coulomb glass value at $W=1$,
and found $\xi^{\SG}_L \leq 1$ as $T\to 0$, which suggests
that the large CL is due to the long-range interaction.

{\em Coulomb gap --}
The single-particle DOS is defined as
$g_L(\e,T)= N^{-1} [\la  \sum_{i=1}^N \delta(\e-\e_i)\ra ]_{av}$ 
where $\e_i = \sum_{j\neq i} (n_j -K)/r_{ij} +W \varphi_i$
 is the cost of adding an electron at site $i$ of
the central cell while leaving the
periodic images unchanged. We compute the infinite sum with
the Ewald method.
Because of the dipole term \cite{leeuw}, 
the DOS has no hard gap 
\cite{dipole,gp}, unlike for a finite, nonperiodic system.
The finite-size effects due to the energy scale $L^{-1}$ 
turn out to be significant 
for $|\e| \leq a L^{-1}$ with $a\simeq 0.3$, while
those due to the sample 
fluctuations of $\mu$ (of order $W/L^{d/2}$) were drastically
reduced by shifting the DOS before averaging over the samples \cite{shift}.
In the gap region $(|\e|, T) \ll \Delta$,
which is our only focus here, one expects the scaling 
$g_L(\e,T)=T^\delta f(|\e|/T)$ for $a L^{-1} \ll (|\e|, T)$, with
  $f(x)\sim$ constant as $x\to 0$ and $f(x)\sim c\, x^\delta$ as $x\to \infty$.  
Fig.~4(a,b,c) show scaling plots with $\delta=2$ 
for $L=10$ and $W=4,2,0.5$.
For $W=4$ scaling is excellent even for 
this moderate size, with small deviations
for $|\e| \lesssim 0.03$ due to finite-size effects.
For $\Delta/T \gg |\e|/T \gtrsim 6$ the data
are well fitted by $g_{10}(\e,T) = c |\e|^2$  with 
$c\simeq 1.1$, which is close to the self-consistent prediction $c=3/\pi$ 
\cite{efros} (while Ref.\cite{mueller}
finds $c=0.2083$). 
As shown in Fig.~4(b) (inset),  the finite-size scaling ansatz
$g_L(\e,T)=L^{-\delta} h(\e L)$, 
which should hold for $T\ll |\e| \ll a L^{-1} \ll \Delta$ 
[with $h(x)\sim c |x|^\delta$ for large $x$],
is also well satisfied with
$c=3/\pi$, $\delta=2$. Our final estimate is $\delta=2.01\pm 0.05$,
which provides strong support for a saturated ES bound.

For decreasing $W$, we observe increasingly stronger deviations from
the $\delta=2$ scaling.
A fit $g_{10}(\e,T) = c |\e|^\delta$ at low $T$ 
gives an effective exponent $\delta\simeq 2.3$ for $W=2$ and
$\delta \geq 2.8$ for $W=0.5$ [Fig.~4(c)].
We interpret this as a crossover due 
to the vicinity of the fluid-COP boundary, below which the DOS has 
a hard gap at $T=0$.
The crossover is apparent in
Fig.~4(d): Since $g_L(\e=0,T)/T^2 \propto T^{\delta-2}$ for
$a L^{-1} \leq T \leq \Delta$, the plateau 
for $W=4$ supports $\delta=2$, 
while for decreasing $W$ the exponent increases to $\delta > 3$.
The $L$ dependence  is 
consistent with the scaling $g_L(\e=0,T)=T^\delta h(T L)$
(not shown) with $\delta$ extracted from Fig.4(d)
for each value of $W$.
Our results differ markedly from Ref.\cite{surer}, which reports 
$\delta = 1.83(3)$  for the same model at $W=0.4$, $T=0$
[however, $g_L(\e\simeq 0)$ is much larger than our data, 
and {\em increases} with $L$]. 
A similar crossover in the DOS was reported for $d=2$,
where the COP occurs at $W=0$  \cite{pikus}.

\begin{figure}

    \begin{tabular*}{\linewidth}{@{\extracolsep{\fill}} l r}

        \includegraphics[height=0.55\linewidth,angle=270]{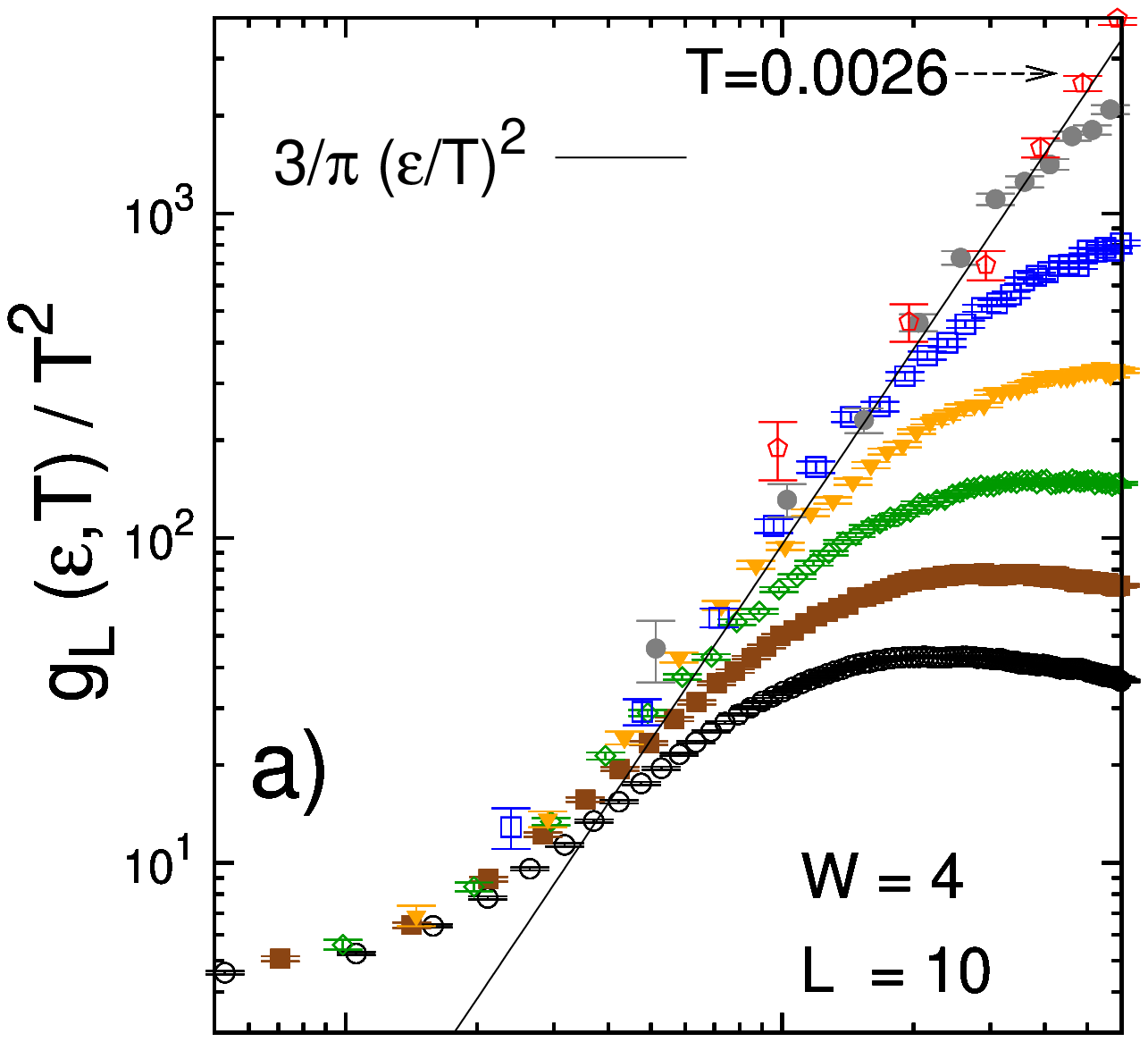} &
        \vspace{-0.35cm}
        \includegraphics[height=0.43\linewidth,angle=270]{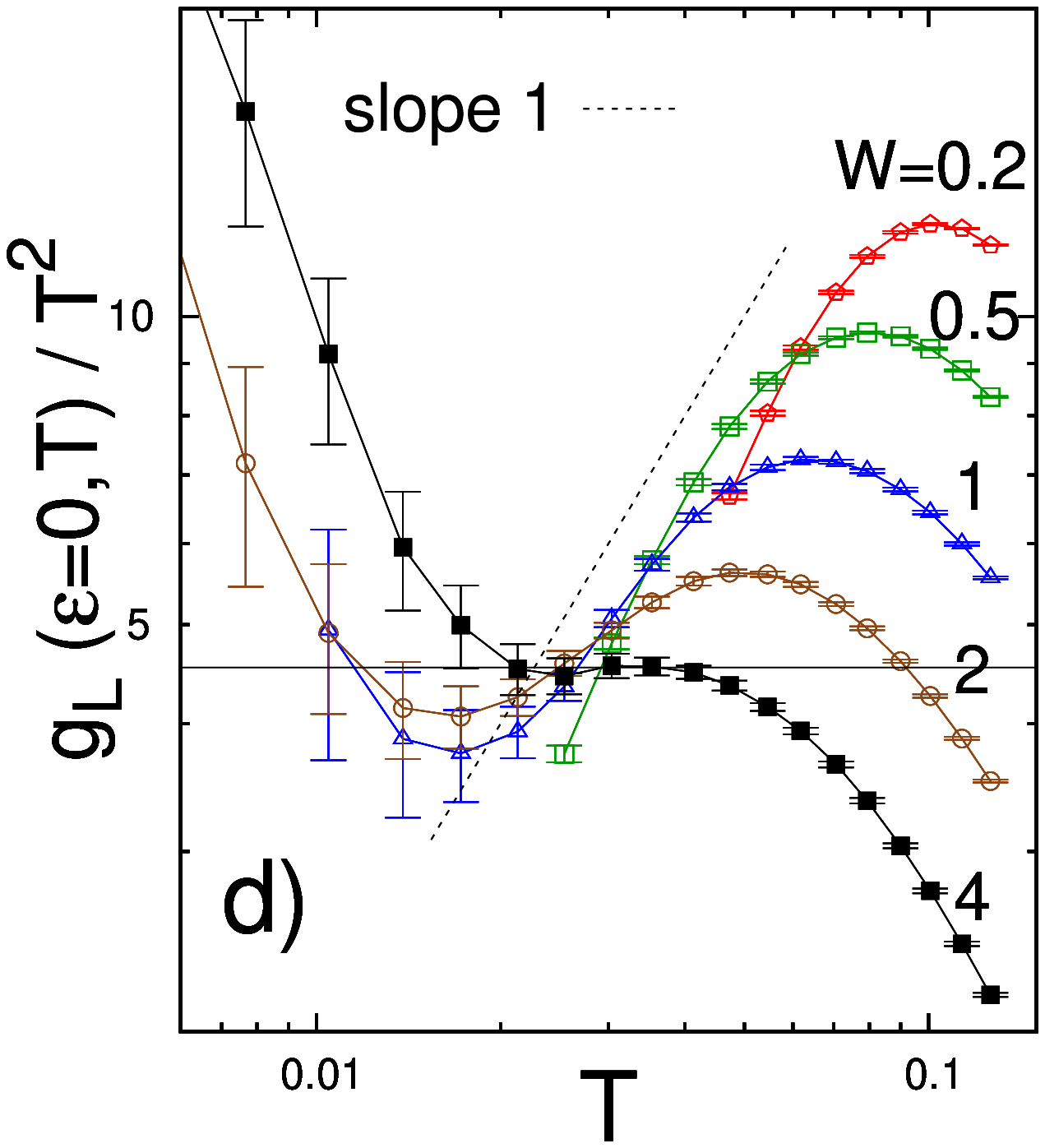} \\

        \includegraphics[height=0.55\linewidth,angle=270]{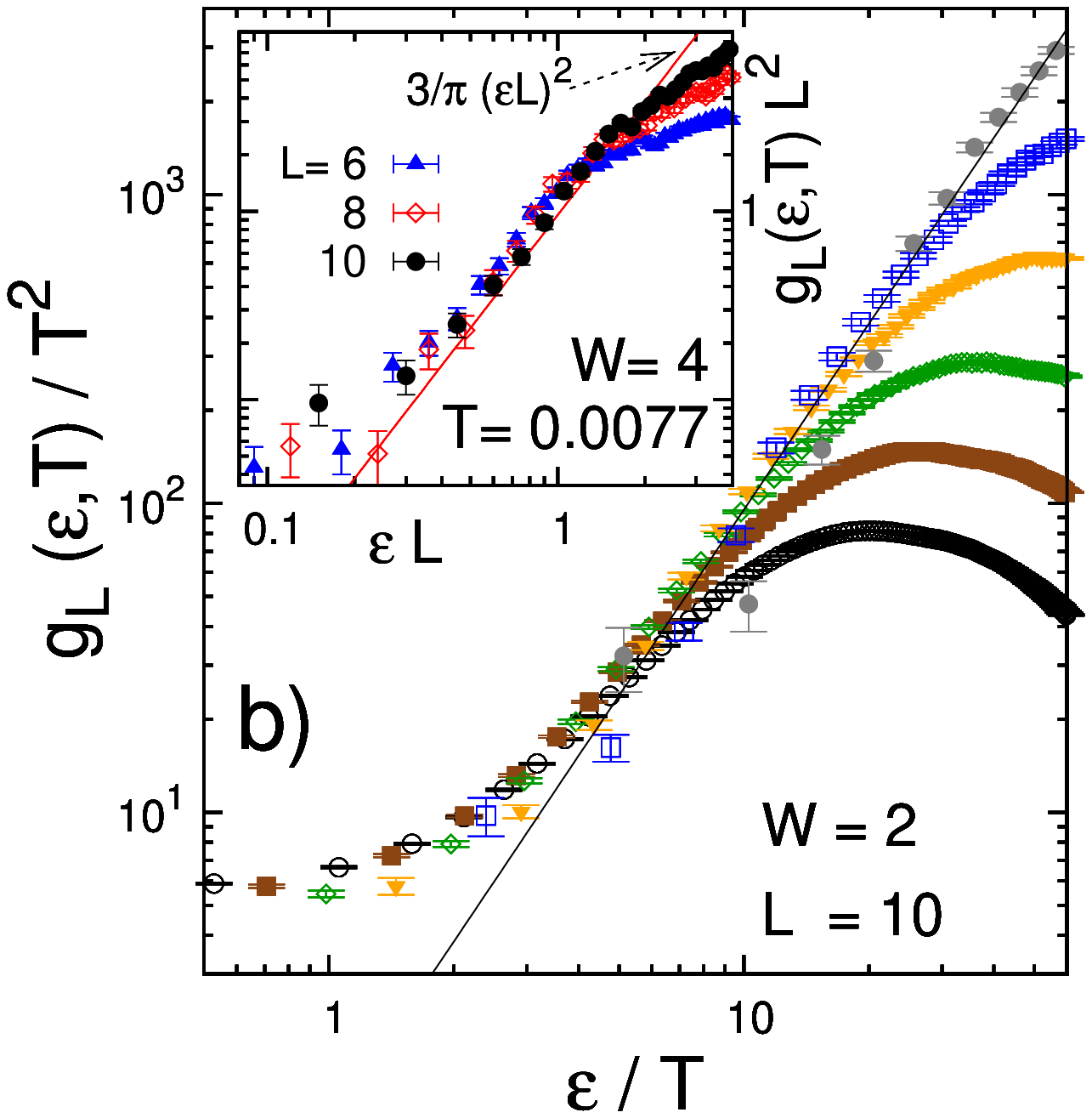} &
\hspace{-0.248cm}    
\includegraphics[height=0.44597\linewidth,angle=270]{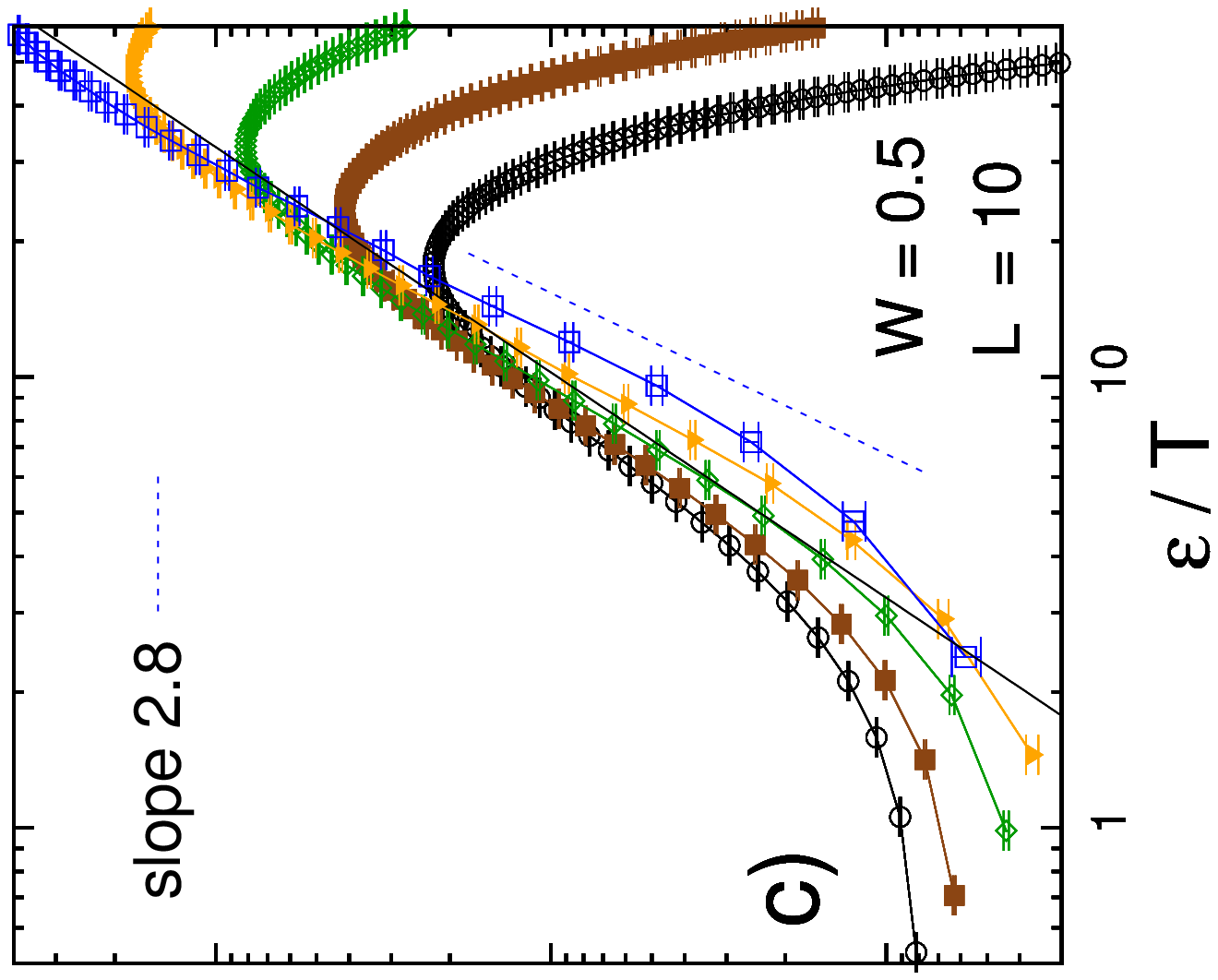} \\
    \end{tabular*}
\caption{(Color online) {\bf (a,b,c)} Scaling plots 
of the DOS for $W=4, 2, 0.5$  and $L=10$. 
From bottom to top: $T = 0.047,0.035,0.025,0.017,0.0105$
plus $T=0.0049$ for $W=4,2$ and $T=0.0026$ for $W=4$.  
The error is the
standard deviation of the sample fluctuations. The 
solid lines represent the ES law $g(\e,T)=3 \,\e^2/\pi$.
The dashed line with slope 2.8 in {\bf (c)} highlights 
the departure from the ES law at low $W$ (the 
$W=0.5,T=0.0105$ data are not fully equilibrated,
but the slope {\em increases} with simulation time).
{\bf Inset of (b)} Finite-size scaling for $W=4$ and $T=0.0077$
(other values of $W$ and $T$ give similar plots).
{\bf (d)} Temperature dependence of the DOS at $|\e|\leq 0.0075$ for $L=10$.}
\vspace{-0.3cm}
\end{figure}

In conclusion, we presented evidence that no equilibrium glass phase
exists in the Coulomb glass, but a saturated ES bound holds. 
The long-range part of the interaction appears to be 
irrelevant as to the equilibrium thermodynamics, except for a possible diverging 
correlation length at $T=0$, which calls for further investigation.

We thank M.~M\"uller, V.~Dobrosavljevic, A.L.~Efros, H.~Katzgraber, A.~M\"obius, 
and G.~Zimanyi for discussions.
This work is supported by the Generalitat de
Catalunya and the Ministerio de Ciencia e Innovaci\'on 
(FIS-2006-13321-C02-01, AP2007-01005).
The computations were performed on 
the BSC-RES node at Universidad de Cantabria and the Albeniz cluster at UB.
MP thanks the Aspen Center for Physics for hospitality.


\vspace{-0.6cm}




\begin{thebibliography}{68}
\expandafter\ifx\csname natexlab\endcsname\relax\def\natexlab#1{#1}\fi
\expandafter\ifx\csname bibnamefont\endcsname\relax
  \def\bibnamefont#1{#1}\fi
\expandafter\ifx\csname bibfnamefont\endcsname\relax
  \def\bibfnamefont#1{#1}\fi
\expandafter\ifx\csname citenamefont\endcsname\relax
  \def\citenamefont#1{#1}\fi
\expandafter\ifx\csname url\endcsname\relax
  \def\url#1{\texttt{#1}}\fi
\expandafter\ifx\csname urlprefix\endcsname\relax\def\urlprefix{URL }\fi
\providecommand{\bibinfo}[2]{#2}
\providecommand{\eprint}[2][]{\url{#2}}



\bibitem{ES}
\bibinfo{author}{\bibfnamefont{A. L.}~\bibnamefont{Efros}} \bibnamefont{and}
  \bibinfo{author}{\bibfnamefont{B. I.}~\bibnamefont{Shklovskii}},
  \bibinfo{journal}{J. Phys. C} \textbf{\bibinfo{volume}{8}},
  \bibinfo{pages}{L49} (\bibinfo{year}{1975}).

\bibitem{crossover} See e.g. 
A. M{\"o}bius, J. Phys. C {\bf 18}, 4639 (1985); 
A.G. Zabrodskii, Phil. Mag. B {\bf 81}, 1131 (2001).



\bibitem{li_phillips}
\bibinfo{author}{\bibfnamefont{Q.}~\bibnamefont{Li}} \bibnamefont{and}
  \bibinfo{author}{\bibfnamefont{P.}~\bibnamefont{Phillips}},
  \bibinfo{journal}{Phys. Rev. B} \textbf{\bibinfo{volume}{49}},
  \bibinfo{pages}{10269} (\bibinfo{year}{1994}).




\bibitem[{\citenamefont{{M{\"o}bius} et~al.}(1992)\citenamefont{{M{\"o}bius},
  {Richter}, and {Drittler}}}]{moebius_dos}
\bibinfo{author}{\bibfnamefont{A.}~\bibnamefont{{M{\"o}bius}}},
  \bibinfo{author}{\bibfnamefont{M.}~\bibnamefont{{Richter}}},
  \bibnamefont{and}
  \bibinfo{author}{\bibfnamefont{B.}~\bibnamefont{{Drittler}}},
  \bibinfo{journal}{Phys. Rev. B} \textbf{\bibinfo{volume}{45}},
  \bibinfo{pages}{11568} (\bibinfo{year}{1992}).


\bibitem{sarvestani}
\bibinfo{author}{\bibfnamefont{M.}~\bibnamefont{Sarvestani}},
  \bibinfo{author}{\bibfnamefont{M.}~\bibnamefont{Schreiber}}, \bibnamefont{and}
  \bibinfo{author}{\bibfnamefont{T.}~\bibnamefont{Vojta}},
  \bibinfo{journal}{Phys. Rev. B} \textbf{\bibinfo{volume}{52}},
  \bibinfo{pages}{R3820} (\bibinfo{year}{1995}).



\bibitem{overlin}
\bibinfo{author}{\bibfnamefont{M.H.}~\bibnamefont{Overlin}},
  \bibinfo{author}
{\bibfnamefont{L.A.}~\bibnamefont{Wong}}, 
  \bibinfo{author}
\bibnamefont{and}
  \bibinfo{author}{\bibfnamefont{C. C.}~\bibnamefont{Yu}},
  \bibinfo{journal}{Phys. Rev. B} \textbf{\bibinfo{volume}{70}},
  \bibinfo{pages}{214203} (\bibinfo{year}{2004}).

\bibitem{surer}
\bibinfo{author}{\bibfnamefont{B.}~\bibnamefont{Surer}},
  \bibinfo{author}{\bibfnamefont{H. G.}~\bibnamefont{Katzgraber}},
  \bibinfo{author}{\bibfnamefont{G. T.}~\bibnamefont{Zimanyi}},
  \bibinfo{author}{\bibfnamefont{B. A.}~\bibnamefont{Allgood}}, \bibnamefont{and}  
  \bibinfo{author}{\bibfnamefont{G.}~\bibnamefont{Blatter}},
  \bibinfo{journal}{Phys. Rev. Lett.} \textbf{\bibinfo{volume}{102}},
  \bibinfo{pages}{067205}(\bibinfo{year}{2009}).

\bibitem{levin}
\bibinfo{author}{\bibfnamefont{E. I.}~\bibnamefont{Levin}},
  \bibinfo{author}{\bibfnamefont{V. L.}~\bibnamefont{Nguyen}},
  \bibinfo{author}{\bibfnamefont{B. I.}~\bibnamefont{Shklovskii}}, \bibnamefont{and}
  \bibinfo{author}{\bibfnamefont{A. L.}~\bibnamefont{Efros}},
  \bibinfo{journal}{Sov. Phys. JETP} \textbf{\bibinfo{volume}{66}},
  \bibinfo{pages}{842} (\bibinfo{year}{1987}).





\bibitem{vjs}
\bibinfo{author}{\bibfnamefont{T.}~\bibnamefont{Vojta}},
  \bibinfo{author}{\bibfnamefont{W.}~\bibnamefont{John}}, \bibnamefont{and}
  \bibinfo{author}{\bibfnamefont{M.}~\bibnamefont{Schreiber}},
  \bibinfo{journal}{J. Phys. Condens. Matter} \textbf{\bibinfo{volume}{5}},
  \bibinfo{pages}{4989} (\bibinfo{year}{1993}).



\bibitem{hunt}
\bibinfo{author}{\bibfnamefont{A.}~\bibnamefont{Hunt}},
  \bibinfo{journal}{Philos. Mag. Lett.} \textbf{\bibinfo{volume}{62}},
  \bibinfo{pages}{371} (\bibinfo{year}{1990}).


\bibitem{davies}
\bibinfo{author}{\bibfnamefont{J. H.}~\bibnamefont{Davies}},
\bibinfo{author}{\bibfnamefont{P. A.}~\bibnamefont{Lee}}, \bibnamefont{and}
  \bibinfo{author}{\bibfnamefont{T. M.}~\bibnamefont{Rice}},
  \bibinfo{journal}{Phys. Rev. B} \textbf{\bibinfo{volume}{29}},
  \bibinfo{pages}{4260} (\bibinfo{year}{1984}).

\bibitem{see} See e.g. Ref.\cite{mueller} and references therein.

\bibitem{mueller}
\bibinfo{author}{\bibfnamefont{M.}~\bibnamefont{M{\"u}ller}} \bibnamefont{and}
  \bibinfo{author}{\bibfnamefont{S.} \bibnamefont{Pankov}},
  \bibinfo{journal}{Phys. Rev. B} \textbf{\bibinfo{volume}{75}},
  \bibinfo{pages}{144201} (\bibinfo{year}{2007}).



\bibitem{grannan}
\bibinfo{author}{\bibfnamefont{E. R.}~\bibnamefont{Grannan}} \bibnamefont{and}
  \bibinfo{author}{\bibfnamefont{C. C.}~\bibnamefont{Yu}},
  \bibinfo{journal}{Phys. Rev. Lett.} \textbf{\bibinfo{volume}{71}},
  \bibinfo{pages}{3335} (\bibinfo{year}{1993}).



\bibitem{diaz}
\bibinfo{author}{\bibfnamefont{A.}~\bibnamefont{D\'iaz-S\'anchez}},
  \bibinfo{author}{\bibfnamefont{M.}~\bibnamefont{Ortu\~no}},
  \bibinfo{author}{\bibfnamefont{A.}~\bibnamefont{P\'erez-Garrido}}, \bibnamefont{and}
  \bibinfo{author}{\bibfnamefont{E.}~\bibnamefont{Cuevas}},
  \bibinfo{journal}{Phys. Stat. Sol. (b)} \textbf{\bibinfo{volume}{218}},
  \bibinfo{pages}{11} (\bibinfo{year}{2000}).


\bibitem{AT} J.R.L. de Almeida and D.J.~Thouless, J. Phys. A {\bf 11}, 983 
(1978).


\bibitem{pastor}
\bibinfo{author}{\bibfnamefont{A. A.}~\bibnamefont{Pastor}} \bibnamefont{and}
  \bibinfo{author}{\bibfnamefont{V.}~\bibnamefont{Dobrosavljevi{\'c}}},
  \bibinfo{journal}{Phys. Rev. Lett.} \textbf{\bibinfo{volume}{83}},
  \bibinfo{pages}{4642} (\bibinfo{year}{1999}).



\bibitem[{\citenamefont{Pankov and Dobrosavljevi{\'c}}(2005)}]{pankov}
\bibinfo{author}{\bibfnamefont{S.}~\bibnamefont{Pankov}} \bibnamefont{and}
  \bibinfo{author}{\bibfnamefont{V.}~\bibnamefont{Dobrosavljevi{\'c}}},
  \bibinfo{journal}{Phys. Rev. Lett.} \textbf{\bibinfo{volume}{94}},
  \bibinfo{pages}{046402} (\bibinfo{year}{2005}).


\bibitem[{\citenamefont{M{\"u}ller and Ioffe}(2004)}]{ioffe}
\bibinfo{author}{\bibfnamefont{M.}~\bibnamefont{M{\"u}ller}} \bibnamefont{and}
  \bibinfo{author}{\bibfnamefont{L.~B.} \bibnamefont{Ioffe}},
  \bibinfo{journal}{Phys. Rev. Lett.} \textbf{\bibinfo{volume}{93}},
  \bibinfo{pages}{256403} (\bibinfo{year}{2004}).


\bibitem{efros}
A. L. Efros, J. Phys. C {\bf 9}, 2021 (1976).



\bibitem{moebius} A.~M{\"obius} and U. K.~R{\"o}{\ss}ler, arxiv:0904.3723 (2009).




\bibitem{malik}
\bibinfo{author}{\bibfnamefont{V.}~\bibnamefont{Malik}}
  \bibinfo{author}
\bibnamefont{and}
  \bibinfo{author}{\bibfnamefont{D.}~\bibnamefont{Kumar}},
  \bibinfo{journal}{Phys. Rev. B} \textbf{\bibinfo{volume}{76}},
  \bibinfo{pages}{125207} (\bibinfo{year}{2007}).


\bibitem{moebius_talk}
A. M\"obius, talk given at TIDS11 (2005).



\bibitem{gp} M.~Goethe and M.~Palassini, in preparation.

\bibitem{leeuw}
S. W. de Leeuw, J. W. Perram,  E. R. Smith, 
Proc. R. Soc.    London, Ser. A {\bf 373}, 27 (1980).



\bibitem{exchange}
\bibinfo{author}{\bibfnamefont{K.}~\bibnamefont{Hukushima}}
  \bibnamefont{and}
  \bibinfo{author}{\bibfnamefont{K.}~\bibnamefont{Nemoto}},
  \bibinfo{journal}{J. Phys. Soc. Japan} \textbf{\bibinfo{volume}{65}},
  \bibinfo{pages}{1604} (\bibinfo{year}{1996}).


\bibitem{method}
The largest
size simulated is $L =(10, 8, 10)$
and the number of $(T,W)$ pairs is $(91, 67, 37)$,
for (ABC, ADE, $W=$ constant), respectively.
The number of samples
is between 100 and 676 for $L=10$, depending on the path, and
larger for $L<10$. For each triplet $(\varphi,T,W)$, we simulate
four independent replicas to obtain an unbiased estimate
of $\xi^{\SG}_{\tiny{L}}$.
We used $t_{s}=10^5$ MC sweeps per replica for paths ABC 
and $W=\text{constant}$,
and $t_{s}=3 \cdot 10^5$ for ADE, for a total of about $2.5 \times 10^5$ computing hours.


\bibitem{cooper} F.~Cooper, B.~Freedman, and D.~Preston, 
Nucl. Phys. B {\bf 210}, 210 (1982).



\bibitem{ballesteros2}
H. G. Ballesteros et al., Phys. Rev. B {\bf 62}, 14237 (2000);
M. Palassini and S. Caracciolo, Phys. Rev. Lett. {\bf 82}, 5128 (1999).

\bibitem{fisher} See e.g. D. S. Fisher, Phys. Rev. Lett. {\bf 56}, 416 (1986).

\bibitem{ballesteros}
\bibinfo{author}{\bibfnamefont{H. G.}~\bibnamefont{Ballesteros}},
  \bibinfo{author}{\bibfnamefont{L. A.}~\bibnamefont{Fern\'andez}},
  \bibinfo{author}{\bibfnamefont{V.}~\bibnamefont{Mart\'in-Mayor}}, \bibnamefont{and}
  \bibinfo{author}{\bibfnamefont{A.}~\bibnamefont{Mu\~noz Sudupe}},
  \bibinfo{journal}{Phys. Lett. B} \textbf{\bibinfo{volume}{378}},
  \bibinfo{pages}{207} (\bibinfo{year}{1996}).%



\bibitem{middleton}
\bibinfo{author}{\bibfnamefont{A. A.}~\bibnamefont{Middleton}}
  \bibinfo{author}
\bibnamefont{and}
  \bibinfo{author}{\bibfnamefont{D. S.}~\bibnamefont{Fisher}},
  \bibinfo{journal}{Phys. Rev. B} \textbf{\bibinfo{volume}{65}},
  \bibinfo{pages}{134411} (\bibinfo{year}{2002}).


\bibitem{tf}
A naive application of the Thomas-Fermi theory
gives a screening length diverging as $g(\e=0,T)^{-1/2}\sim T^{-\delta/2}$.


\bibitem{dipole}
The energy change for an electron hop from $i$ to $j$ in all 
images receives a positive contribution $4 \pi |{\bf r}_{ij}|^2 / (3 L^3)$.




\bibitem{shift}
For each sample we shift
$\e$ by $(\e_a + \e_b)/2$, where
$\int_{-\infty}^{\e_a} d\e \,
g_L(\e,T,\varphi) =\int_{\e_b}^{\infty} d \e \,
g_L(\e,T,\varphi) = 0.499$. 


\bibitem{pikus} F.G.~Pikus and A.L.~Efros, Phys. Rev. Lett. {\bf 73}, 3014
(1994).












\end{thebibliography}
\end{document}